%% file: ms.tex
\documentclass{article}

\usepackage{arxiv}

\usepackage[utf8]{inputenc} 
\usepackage[T1]{fontenc}    
\usepackage{hyperref}       
\usepackage{url}            
\usepackage{booktabs}       
\usepackage{amsfonts}       
\usepackage{nicefrac}       
\usepackage{microtype}      
\usepackage{graphicx}
\usepackage{natbib}
\usepackage{doi}
\usepackage{xcolor}
\usepackage{enumerate}
\usepackage{amsmath}
\usepackage{braket}
\usepackage{subcaption}

\usepackage{tikz}
\usepackage{algorithm}
\usepackage{algorithmic}
\usepackage{siunitx}

\usepackage{draftwatermark}
\SetWatermarkText{}

\newcommand{\R}{\mathbb{R}}
\newcommand{\Q}{\mathbb{E}}
\newcommand*\diff{\mathop{}\!\mathrm{d}}
\newcommand{\bs}{\boldsymbol}
\newcommand{\mc}{\mathcal}

\newcommand{\smallminus}{\scalebox{1}[1]{-}}

\title{Quantum-Inspired Solver for \\ Simulating Material Deformations}

\author{ Mazen Ali \\
	Multiverse Computing\\
	Paseo de Miramón\\
	E-20014 San Sebastián, Spain \\
	\texttt{mazen.ali@multiversecomputing.com} \\
    \And
	Aser Cortines \\
	Multiverse Computing\\
	Paris-Saclay\\ 
    91120 Palaiseau, France\\
	\texttt{aser.cortines@multiversecomputing.com} \\
    \And
	\hspace{1mm}Siddhartha Morales \\
	Multiverse Computing\\
	Paseo de Miramón\\
	E-20014 San Sebastián, Spain \\
	\texttt{siddhartha.morales@multiversecomputing.com} \\
    \And
	\hspace{1mm} Samuel Mugel \\
	Multiverse Computing \\
    192 Spadina Ave \\
    509 Toronto, Canada \\
	\texttt{sam.mugel@multiversecomputing.com} \\
    \And
	\hspace{1mm} Mireia Olave \\
	Ikerlan Technology Research Centre, \\
    Jose Maria Arizmendiarreta pasealekua, 2 \\
    20500 Mondragon, Spain \\
	\texttt{molave@ikerlan.es} \\
    \And
	\hspace{1mm}Roman Orus \\
	Multiverse Computing, Paseo de Miramón E-20014 San Sebastián, Spain \\
    Donostia International Physics Center, Paseo Manuel de Lardizabal 4, E-20018 San Sebastián, Spain \\
    Ikerbasque Foundation for Science, Maria Diaz de Haro 3, E-48013 Bilbao, Spain \\
	\texttt{roman.orus@multiversecomputing.com} \\
    \And
    \hspace{1mm}Samuel Palmer \\
	Multiverse Computing \\
    192 Spadina Ave \\
    509 Toronto, Canada \\
	\texttt{sam.palmer@multiversecomputing.com}  \\
    \And
	\hspace{1mm} Hodei Usabiaga \\
	Ikerlan Technology Research Centre, \\
    Jose Maria Arizmendiarreta pasealekua, 2 \\
    20500 Mondragon, Spain \\
	\texttt{husabiaga@ikerlan.es} \\
}

\renewcommand{\headeright}{}
\renewcommand{\undertitle}{}
\renewcommand{\shorttitle}{TN-FEM}

\hypersetup{
pdftitle={TN-FEM Elasticity Solver},
pdfsubject={quanth-ph, math.NA},
pdfauthor={Mazen Ali},
pdfkeywords={Elasticity Tensor Networks Solver},
}

\begin{document}

\maketitle

\renewcommand{\thefootnote}{\fnsymbol{footnote}}
\footnotetext[1]{The authors are listed in alphabetical order.}
\renewcommand{\thefootnote}{\arabic{footnote}}

\begin{abstract}
This paper explores the application of tensor networks (TNs) to the simulation of material deformations within the framework of linear elasticity. Material simulations are essential computational tools extensively used in both academic research and industrial applications. TNs, originally developed in quantum mechanics, have recently shown promise in solving partial differential equations (PDEs) due to their potential for exponential speedups over classical algorithms. Our study successfully employs TNs to solve linear elasticity equations with billions of degrees of freedom, achieving exponential reductions in both memory usage and computational time. These results demonstrate the practical viability of TNs as a powerful classical backend for executing quantum-inspired algorithms with significant efficiency gains. This work is based on our research conducted with IKERLAN.
\end{abstract}

\keywords{Quantum-Inspired \and Tensor Networks \and Partial Differential Equations
\and Linear Elasticity \and Constitutive Relations \and Material Deformations \and Finite Element Methods}

\section{Introduction}

Numerical simulations have become a cornerstone in material science, particularly in the field of solid mechanics, where they are used to simulate stresses in mechanical structures. These simulations allow researchers and engineers to predict the behavior of materials under various conditions, which is crucial for both scientific understanding and industrial applications. The ability to model material behavior accurately helps in designing materials and structures that are more efficient, durable, and safe.

In the context of solid mechanics, the finite element method (FEM) is the predominant numerical technique used for simulating deformations, see \cite{braess07}.
The method divides a complex geometry into smaller, manageable pieces called elements, and applies equations that describe the material behavior to these elements. This method is particularly powerful because it can handle complex shapes, boundary conditions, and material properties, making it suitable for a wide range of applications from small-scale material analysis to large-scale structural simulations.

One of the primary challenges associated with FEM is the significant computational resources it requires, particularly for simulations involving complex geometries, fine meshes, or highly nonlinear material behaviors. As the complexity of the problem increases, the time and memory demands can become prohibitive, limiting the practicality of FEM for large-scale or real-time applications.

In light of these constraints, there is growing interest in exploring alternative approaches that can offer more efficient solutions. One promising avenue is the use of quantum-inspired methods for partial differential equations,
see \cite{bachmayr2023}.
These methods leverage principles derived from the simulation of
quantum systems to enhance computational efficiency and overcome some of the limitations inherent in classical numerical techniques.

The quantum-inspired technique utilized in our research is tensor networks (TNs). Tensor networks are mathematical structures that efficiently represent high-dimensional data through a network of interconnected tensors, see \cite{orus2014, biamonte17}.
In the context of material simulations, TNs offer a powerful framework for handling the complex interactions present in large systems. By decomposing the system into smaller, more manageable components, TNs can significantly reduce the computational complexity and memory requirements compared to traditional methods like FEM.
The ability of TNs to capture intricate details with improved efficiency positions them as a valuable tool in advancing the field of material science.

\subsection{Previous Work}
Numerous studies have explored the applications of TNs across various fields including physics, chemistry, optimization, finance, and machine learning. Most relevant to this study are the works in \cite{markeeva21, bachmayr20, kornev23}.
In \cite{markeeva21}, the authors implemented a TN solver for the Poisson problem, which serves as our primary reference for this study.
The authors demonstrated that the memory usage and time to solution of the PDE solver scale only logarithmically with respect to grid size, in contrast to the at least linear scaling observed with classical solvers.
Building upon this,
\cite{kornev23, kornev24} implemented a 2D Navier-Stokes solver and demonstrated similar speedups.
Additionally, \cite{bachmayr20} describes a preconditioner for the Poisson problem, which is pertinent to elasticity problems as well.
The authors demonstrate that, with a specialized preconditioner for the TT format, TT-based PDE solvers maintain stability for grid resolutions that exceed numerical precision.

In the process of submitting this work,
we became aware of a similar preprint that appeared a week prior -- see
\cite{benvenuti2025}.

\subsection{This Work}
This work is dedicated to developing a numerical TN-FEM solver for elasticity equations,
first of its kind.
In Section \ref{sec:linelast}, we provide a brief introduction to linear elasticity.
In Section \ref{sec:fem}, we introduce FEM
and, in Section \ref{sec:tns}, TNs.
In Section \ref{sec:methods}, we discuss some aspects of the implementation.
In Section \ref{sec:results}, we demonstrate the performance of the TN solver
and compare it to a classical FEM solver.
Finally, in Section \ref{sec:concl}, we summarize and provide
an outlook to future work.

\section{Linear Elasticity}\label{sec:linelast}
The equations of elasticity in three dimensions are formulated through constitutive relations between infinitesimal strains and stresses in a material. For small deformations, these relationships can be assumed to be linear.

Let $\Omega \subset \R^3$ be a domain representing a material solid. Displacements are modeled with a vector field \( u:\Omega \rightarrow \R^3 \). The infinitesimal strain at a point \( x \in \Omega \) in a material undergoing the deformation \( u \) is quantified by the \emph{strain tensor}
\begin{equation*}
    \varepsilon(x) := \varepsilon(u(x)) := \frac{1}{2} (\nabla u(x) + (\nabla u(x))^T).
\end{equation*}
The \emph{Cauchy stress tensor}
\begin{equation*}
\sigma(x) = \begin{pmatrix}
\sigma_{xx}(x) & \sigma_{xy}(x) & \sigma_{xz}(x) \\
\sigma_{yx}(x) & \sigma_{yy}(x) & \sigma_{yz}(x) \\
\sigma_{zx}(x) & \sigma_{zy}(x) & \sigma_{zz}(x)
\end{pmatrix}
\end{equation*}
quantifies all infinitesimal stress forces within a material. Conservation laws impose that this tensor is symmetric, having only 6 independent components.

In linear elasticity (small displacements and small stresses), the relationship between stresses and strains is fully defined through a fourth-order \emph{material stiffness tensor}
\( C(x) \in \R^{3 \times 3 \times 3 \times 3} \):
\begin{equation}\label{eq:linelastic}
    \sigma(x) = C(x) : \varepsilon(x),
\end{equation}
where “$:$” denotes the tensor double-dot product. This relationship can be greatly simplified by assuming the material is homogeneous (\( C(x) \equiv C \)) and isotropic (material properties are independent of coordinate direction). In this case, only two scalar parameters (the \emph{Lamé coefficients}), \(\lambda\) and \(\mu\), are needed, and \eqref{eq:linelastic} can be restated as
\begin{equation*}
    \sigma = \lambda \mathrm{tr}(\varepsilon) I_{3 \times 3} + 2 \mu \varepsilon,
\end{equation*}
where \( x \) is omitted for brevity.

Finally, for a given applied external force \( f : \Omega \rightarrow \R^3 \), the change in stresses satisfies
\begin{equation*}
    \nabla \cdot \sigma = f.
\end{equation*}
Altogether, we have the differential equation
\begin{align}\label{eq:diffeq}
    \nabla \cdot \sigma &= f, \notag\\
    \sigma &= \lambda \mathrm{tr}(\varepsilon) I_{3 \times 3} + 2 \mu \varepsilon,
\end{align}
with appropriate boundary conditions for the displacement \( u \). This equation can be further simplified by considering plane stresses, such as in a thin metallic plate. In this case, the stresses in the direction perpendicular to the cross-section are negligible, and \eqref{eq:diffeq} reduces to a two-dimensional equation with modified Lamé parameters. This is the scenario considered for the demonstration in Section \ref{sec:results}.

\section{Finite Element Methods}\label{sec:fem}
The standard approach to the numerical solution of \eqref{eq:diffeq} is the \emph{finite element method} (FEM). First, the problem is formulated in a \emph{weak} (integral) form: choosing appropriate functional spaces \( U \) and \( V \), we seek \( u \in U \) such that
\begin{equation*}
    a(u, v) := \int_\Omega \sigma(u(x)) : \varepsilon(v(x)) \, \diff x 
    = L(v) := \int_\Omega f(x) \cdot v(x) \, \diff x,
\end{equation*}
is satisfied for any test function \( v \in V \). Boundary conditions are omitted for simplicity.

The infinite-dimensional functional spaces \( U \) and \( V \) are then replaced by finite-dimensional subspaces \( U_h \subset U \) and \( V_h \subset V \). The computational task is to find \( u_h \in U_h \) such that
\begin{equation}\label{eq:finbilform}
    a(u_h, v_h) = L(v_h), \quad \forall v_h \in V_h.
\end{equation}
The spaces \( U_h \) and \( V_h \) are constructed by discretizing the domain \( \Omega \) into a finite number of cells \( \Omega_m \) (or \emph{elements}) and defining a set of piecewise polynomial basis functions \( \varphi_i \) over those elements. We then set
\begin{equation*}
    U_h := \mathrm{span}\{\varphi_i: i = 0, \ldots, N-1\}.
\end{equation*}
For \emph{Galerkin} methods, we additionally set \( U_h = V_h \). The problem in \eqref{eq:finbilform} can then be restated as a linear system
\begin{equation}\label{eq:femsys}
    \bs A \bs u = \bs f,
\end{equation}
where
\begin{align*}
    \bs u &= (\bs u_i)_i \quad \text{such that} \quad u = \sum_i \bs u_i \varphi_i,\\
    \bs f &= (L(\varphi_i))_i, \quad \text{and} \quad
    \bs A = (a(\varphi_j, \varphi_i))_{i, j}.
\end{align*}
The matrix \( \bs A \) is often referred to as the \emph{stiffness matrix} and \( \bs f \) as the \emph{force vector} or simply the \emph{right-hand side} (RHS).

In classical FEM, before \eqref{eq:femsys} can be solved, we must \emph{assemble} the matrix \( \bs A \) and the RHS \( \bs f \). This is typically done by iterating over elements, computing or approximating the integrals from \eqref{eq:finbilform} over each element, and then adding the result to the \emph{global} stiffness matrix \( \bs A \) and RHS \( \bs f \). To further simplify this computation, basis functions \( \hat{\varphi} \) are defined only over a \emph{reference element}, e.g., in 2D, \( \Q := [-1, 1]^2 \), and then mapped to any element \( \Omega_m \) via a continuously differentiable one-to-one transformation map \( \Phi_m: \Q \rightarrow \Omega_m \).

Let \( x \) indicate coordinates in an element \( \Omega_m \subset \Omega \) and let \( \xi \) indicate coordinates in the reference element \( \Q \). A basis function \( \varphi_i \) on \( \Omega_m \) is defined through a basis function \( \hat{\varphi} \) on \( \Q \) via
\begin{equation*}
    \varphi_i(x) := \hat{\varphi}_i(\Phi_m^{-1}(x)).
\end{equation*}
Then, integrating \eqref{eq:finbilform} over a single element \( \Omega_m \) and abbreviating \( \xi = \Phi_m^{-1}(x) \), we obtain
\begin{equation}\label{eq:elassembly}
    \int_{\Omega_m} \sigma(\varphi_j(x)) : \varepsilon(\varphi_i(x)) \, \diff x
    =
    \int_{\Q} \sigma(\hat{\varphi}_j(\Phi_m^{-1}(x))) : \varepsilon(\hat{\varphi}_i(\Phi_m^{-1}(x))) |\det J_{\Phi_m}(\xi)| \, \diff \xi,
\end{equation}
where \( \det J_{\Phi_m}(\xi) \) is the Jacobian of the transformation \( \Phi_m \) evaluated at \( \xi \). The stress and strain tensors are computed via the gradients of the basis functions, which transform by applying the chain and inverse function rules as
\begin{equation}\label{eq:grads}
    \nabla \varphi_i(x) = \nabla \hat{\varphi}_i(\Phi_m^{-1}(x))
    = J^{-T}_{\Phi_m}(\xi) \nabla \hat{\varphi}_i(\xi).
\end{equation}
The assembly of the RHS is similar.
For more details on FE methods and their application to solid mechanics,
see \cite{braess07}.

\section{Tensor Networks}\label{sec:tns}
\begin{figure}[htbp!]
    \centering
    \input{./images/drawings/qr_dec.tex}
    \caption{QR-decomposition of matrix $A$.}
    \label{fig:qr}
\end{figure}
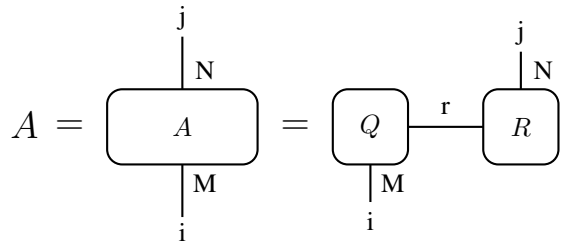

\emph{Tensor networks} (TNs) are a highly efficient representation of higher-order tensors. One of the simplest examples of a TN is the (reduced) QR decomposition:
\begin{equation}\label{eq:qr}
    \bs A = \bs Q \bs R,\quad \bs A \in \R^{M \times N},\; \bs Q \in \R^{M \times r},\; \bs R \in \R^{r \times N}.
\end{equation}
As a \emph{tensor diagram}, we can represent \eqref{eq:qr} as in Figure \ref{fig:qr}

\begin{figure}[htbp!]
    \centering
    \begin{subfigure}[t]{\textwidth}
        \centering
        \input{./images/drawings/TTvec.tex}
        \caption{\textit{Tensor Train} decomposition of a vector, also known as a Matrix Product State.}
        \label{fig:ttvec}
    \end{subfigure}
    \vspace{1em} 
    \begin{subfigure}[b]{\textwidth}
        \centering
        \input{./images/drawings/TTop.tex}
        \caption{\textit{Tensor Train} decomposition of an operator, also known as a Matrix Product Operator.}
        \label{fig:ttop}
    \end{subfigure}
    \caption{Tensor network diagrams.}
    \label{fig:TTdec}
\end{figure}
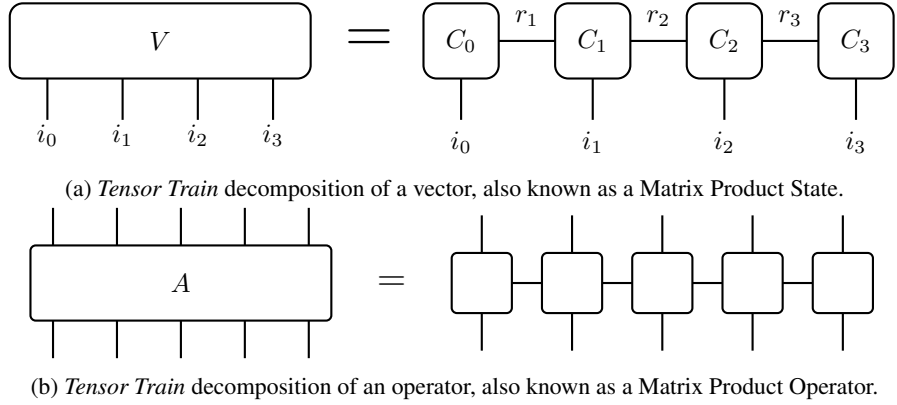

In the above figure, tensors are represented by shapes, indices are represented by lines, and contracted indices correspond to lines shared by tensors. There are numerous other possible TN structures; for a comprehensive introduction,
see \cite{orus2014, biamonte17}. 
For the purposes of this work, \emph{tensor trains} (TTs), also known as Matrix Product States (MPS), as shown in Figure \ref{fig:TTdec} will suffice. More details on the TT decomposition from a mathematical perspective can be found, e.g., in \cite{oseledets11}.

We emphasize that TTs can also be applied to matrices and vectors, a process commonly referred to as the \emph{quantics tensor train} (QTT),
see \cite{oseledets10}.
Any index \(i \in \{0, \ldots, 2^d - 1\}\) can be represented in binary as
\begin{equation*}
    i = (i_0, \ldots, i_{d-1})_2 = \sum_{k=0}^{d-1} i_k 2^{d-1-k}.
\end{equation*}
Thus, we can express a vector \(\bs v \in \R^{2^d}\) of size \(2^d\) as a tensor \(\bs v \in \R^{2 \times \cdots \times 2}\) via
\begin{equation*}
    \bs v(i) = \bs v(i_0, \ldots, i_{d-1}),
\end{equation*}
and apply TN methods to this representation. The same approach can be used for matrices.
In principle, this method can be applied to vectors and matrices of any size by factoring them into primes, but it is simpler to work with sizes that are powers of 2.

The TN compression rate has the potential to provide exponential speedups. For a vector $\bs v \in \R^{N=2^d}$, the explicit storage complexity is of the order $\mc{O}(N)$. In contrast, the storage complexity of an equivalent tensor train (TT) representation is $\mc{O}(r^2 d) = \mc{O}(r^2 \log(N))$, which is only logarithmic in the original size $N$, assuming the ranks $r$ remain small.

\section{Methods: TN-FEM Elasticity Solver}\label{sec:methods}
We demonstrate a solver implementation for the simplest use case described in \eqref{eq:diffeq} in 2D. There are several straightforward and more complex extensions discussed in Section \ref{sec:concl}.
We use \textbf{bold} symbols to indicate vectors, matrices, or tensors. As explained in Section \ref{sec:tns}, matrices and vectors can be treated as tensors with a different representation.

\begin{figure}[htbp!]
    \centering
    \input{./images/drawings/trapezoid.tex}
    \caption{Quadrilateral discretization in powers of 4. As an example we show a $4^2$ discretization, i.e., we have 16 grid points, including boundary points,
    32 d.o.f.\ (16 for each vector field component) and 9 elements.}
    \label{fig:trapezoid}
\end{figure}
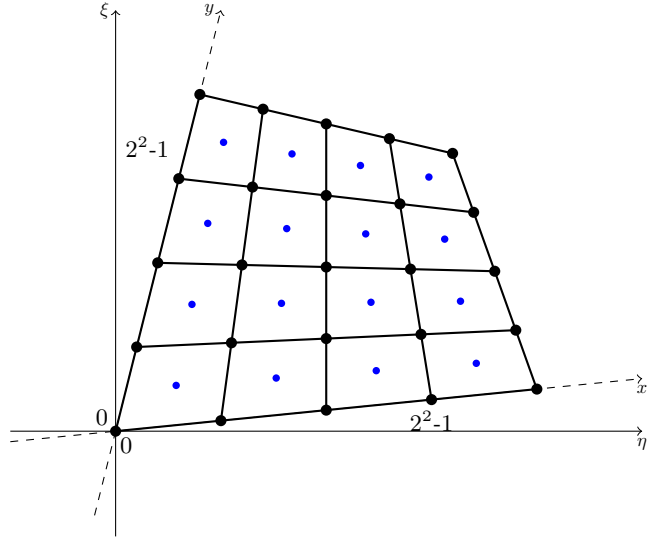

For the domain \(\Omega\), we consider any quadrilateral in \(\R^2\), with either Dirichlet or Neumann boundary conditions on each of the four sides. We discretize this domain as a tensor product grid of \(2^d\) equidistant points along both axes, yielding a total of \(4^d\) discretization points, or \(2\cdot 4^d\) \emph{degrees of freedom} (d.o.f.), and \((2^d - 1)^2\) quadrilateral elements (see Figure \ref{fig:trapezoid}).

We use bilinear finite elements for the basis functions \(\varphi_i\) over this domain to define the trial and test spaces \(U_h\), as detailed in Section \ref{sec:fem}.

To solve \eqref{eq:diffeq}, we need to assemble the system in \eqref{eq:femsys}. The key difference from classical FEM is that all components are now stored and manipulated in a TN format, which requires a specialized assembly process. For a discretization size of \(4^d\), both \(\bs u\) and \(\bs f\) are of size \(2 \cdot 4^d\), i.e., \(4^d\) d.o.f. for each component of the vector fields \(u\) and \(f\). Represented as a TT, \(\bs u\) has the structure like in Figure \ref{fig:ttvec}, and similarly for \(\bs f\). The stiffness matrix \(\bs A\), represented as a TT-operator, has the structure like in Figure \ref{fig:ttop}. Thus, the storage size and computational complexity of all linear algebra operations primarily depend on the ranks of \(\bs u\), \(\bs f\), and \(\bs A\). The cost of manipulating these tensors depends only linearly on \(d\) and, thus, logarithmically on the number of discretization points \(N = 4^d\). The key steps of the assembly process include the computation of Jacobians, the stiffness operator, and the force vector.

\subsection{Jacobians}
The crucial difference between classical FEM assembly and the tensorized assembly in this work lies within the treatment of the Jacobians of the element transformations. Consider again the integral in \eqref{eq:elassembly}. In classical FEM, we iterate over all elements, computing \eqref{eq:elassembly} for each element, and then adding it to the global stiffness matrix $\bs A$. This would involve iterating over all $(2^d - 1)^2$ elements in our case. Instead, we compute \eqref{eq:elassembly} to obtain the entries for all elements $\Omega_m$ in one iteration.

The components in \eqref{eq:elassembly} that depend on the element $\Omega_m$ are the Jacobians $J_{\Phi_m}$, denoted as $J_m$. It was shown in \cite{markeeva21} that for a grid of quadrilaterals, the Jacobians and their determinants at any fixed point $\xi$ are affine linear functions of the element index $m$. Specifically, if the element index $m=(i, j)$ corresponds to the index $i$ along the $x$-axis and $j$ along the $y$-axis, we have
\begin{equation}\label{eq:affinejacs}
    J_{(i, j)}(\xi) =
    J_{(0, 0)}(\xi) + i(J_{(1, 0)}(\xi) - J_{(0, 0)}(\xi))
    + j(J_{(0, 1)}(\xi) - J_{(0, 0)}(\xi)),
\end{equation}
and a similar expression for the determinant. We can then construct two TTs, $\bs X$ and $\bs Y$, such that
\begin{equation*}
    \bs X(i, j) = i \quad \text{and} \quad \bs Y(i, j) = j.
\end{equation*}
This is a tensorized mesh-grid function. The construction of $\bs X$ and $\bs Y$ is straightforward and has a tensor rank of at most 2. Due to \eqref{eq:affinejacs}, since each component of the Jacobian is an affine-linear function in $\bs X$ and $\bs Y$, we can construct a TT for each Jacobian component with the same ranks as $\bs X$ and $\bs Y$. Thus, we obtain
\begin{equation}
\bs J(\xi) = \begin{pmatrix}
\bs J_{11}(\xi) & \bs J_{12}(\xi) \\
\bs J_{21} (\xi)& \bs J_{22}(\xi)
\end{pmatrix},
\end{equation}
where each component $\bs J_{kl}(\xi)$ is a TT representing the values of that component of the Jacobian, evaluated at $\xi \in \Q$, for all elements $m = (i, j)$. The same can be done for the TT $\bs{\det J}$.

All the above computations for $\bs J$ and $\bs{\det J}$ are numerically exact and represented with low-rank TTs. However, the inverse Jacobians in \eqref{eq:grads} have to be expressed explicitly through the components of $J$ to apply TT operations. The inverse of $J$ can be expressed through the adjugate of $J$ and the inverse determinant $(\det J)^{-1}$. In our TT implementation, this means computing the TT representation $(\bs{\det J})^{-1}$ of the component-wise inverse of $\bs{\det J}$. This cannot be done efficiently and exactly in general, but we can use routines such as the \emph{TT-cross} (see \cite{oseledets_cross10})
for an efficient approximation.

\subsection{Stiffness Operator}
Returning to the computation in \eqref{eq:elassembly}, now that we have all the element-dependent components, we need to differentiate between several cases. Each element has four corners, and to each of those corners, we can associate two basis functions (or three in 3D): one for each coordinate direction of the vector fields. Thus, in total, we have eight possible basis functions per element. Since we compute the stiffness operator for pairs of basis functions $a(\varphi_j, \varphi_i)$, this gives a total of 64 possible pairs. Expressing each of these in terms of the tensorized Jacobian components is involved but can be derived and simplified with symbolic math software.

Finally, we need to add each of the terms to the global stiffness operator $\bs A$. Before doing that, we need to determine the correct global position for each of the terms from \eqref{eq:elassembly}. We have TTs such as the ones in
Figure \ref{fig:ttvec}, where,
for a pair of basis functions corresponding to corners $c_1$ and $c_2$,
the TT-vector
$\bs a_{c_1, c_2}(i, j) = \bs a_{c_1, c_2}((i_0, j_0), \ldots, (i_{d-1}, j_{d-1}))$ has $d$ indices, dimension 4 each, and represents the contributions of a specific pair of basis functions to the stiffness operator from all $(2^d - 1)^2$ elements. The global TT-operator $\bs A$ has $d+1$ row and column indices, each of dimensions $2, 4, 4, \ldots$. These do not correspond to the element indices but rather to the global basis function indices.

On a Cartesian grid of quadrilaterals, the conversion between the two is straightforward. It can be performed analogously to \cite{markeeva21}, with adjustments to account for the two vector field components. Finally, boundary conditions can be applied as in \cite{markeeva21}, once again adjusting for vector field components.

\subsection{Force Vector}
For the RHS, the input for the solver is a TT $\bs f^\circ = (\bs f_{j}^\circ)_{j}$ representing the values of $f$ at all the grid points. Since we are using a Lagrange FEM basis, we have the property
\begin{equation*}
    f = \sum_{j} \bs f_{j}^\circ \varphi_{j},
\end{equation*}
i.e., the nodal values are also the linear expansion coefficients of $f$. For the RHS, we thus get
\begin{equation*}
    \bs f = \bs M \bs f^\circ, \quad
    \bs M := \left(\int_\Omega \varphi_j(x) \cdot \varphi_i(x) \diff x \right)_{i, j},
\end{equation*}
where $\bs M$ is the \emph{mass matrix}. The mass operator can be assembled analogously to the stiffness operator. Although, tensor network techniques allow for a direct computation of $ \bs f $ without having to compute $ \bs M$ explicitly.

\subsection{Algorithm}
Having assembled the TT operators $\bs A$, $\bs M$, and the TT vector $\bs f$, we can use various TT solvers for solving $\bs A\bs u = \bs f$. We use the \texttt{AMEn} solver from \cite{amen}. The key steps of the assembly procedure are outlined in Algorithm \ref{alg:assemble}. The assembly of the mass operator $\bs M$ is analogous to Algorithm \ref{alg:assemble}, and the full solver is implemented as in Algorithm \ref{alg:solver}.

\begin{algorithm}
\caption{Assemble Stiffness TN-Operator $\bs A$}
\label{alg:assemble}
\begin{algorithmic}[1]
\REQUIRE Grid size exponent $d$, Young's modulus $E$, Poisson's ratio $\nu$,
quadrilateral $\Omega$,
boundary conditions (Dirichlet or
Neumann on each side of $\Omega$).
\ENSURE TT-operator $\bs A$.
\STATE Initialize TT-operator $\bs A=0$.
\STATE Construct the mesh-grid TTs $\bs X$ and $\bs Y$ as in \cite{markeeva21}.
\STATE Use $\bs X$ and $\bs Y$ to construct $\bs J$ and $\bs\det J$
component-wise, evaluated at $\xi=(0, 0)$.
\STATE Use TT-cross interpolation to approximate the component-wise TT inverse
$(\bs\det J)^{-1}$.
\FOR{each basis function corner pair $(c_1, c_2)$ in local integral \eqref{eq:elassembly} (64 in total)}
    \STATE Compute the TT-operator $\bs A_{c_1, c_2}$ for corner pair
    $(c_1, c_2)$,
    using the material parameters $E$, $\nu$, and the Jacobian terms
    $\bs J$, $\bs\det J$, $(\bs\det J)^{-1}$.
    \STATE Add to global stiffnes $\bs A = \bs A + \bs A_{c_1, c_2}$.
\ENDFOR
\STATE Apply boundary conditions to $\bs A$.
\RETURN $\bs A$
\end{algorithmic}
\end{algorithm}

\begin{algorithm}
\caption{TN-FEM Elasticity Solver}
\label{alg:solver}
\begin{algorithmic}[1]
\REQUIRE Same as for Algorithm \ref{alg:assemble} and
TT-vector $\bs f^\circ$ of values of $f$ on grid points.
\ENSURE TT-vector solution $\bs u$.
\STATE Use Algorithm \ref{alg:assemble} to assemble
stiffness TT-operator $\bs A$ and mass TT-operator $\bs M$.
\STATE Apply \texttt{AMEn} to solve $\bs A\bs u=\bs M\bs f^\circ$.
\RETURN $\bs u$
\end{algorithmic}
\end{algorithm}

\section{Results}\label{sec:results}
\captionsetup[figure]{name={Figure}}
We consider a 2D cantilever beam made of aluminum. The beam has a length of \qty{20}{\metre}, a width of \qty{1}{\metre}, a Young's modulus of $E=\qty{68}{\giga\pascal}$, a Poisson's ratio of $\nu=0.33$, and a density of $\rho=\qty{2700}{\kg\per\metre\cubed}$. The beam is fixed on one side, with only its own weight acting as an external homogeneous force, i.e., \( f = (0, -\rho g) \). We use a grid of quadrilaterals with the reference element $\Q = [-1, 1]^2$ and bi-linear shape functions. For a grid exponent $d$, the number of grid points is $4^d$, the number of elements is $(2^d - 1)^2$, and the number of d.o.f.\ is $2 \cdot 4^d$.

\begin{figure}[htbp!]
    \centering
    \begin{subfigure}[t]{\textwidth}
        \centering
        \includegraphics[width=0.8\textwidth]{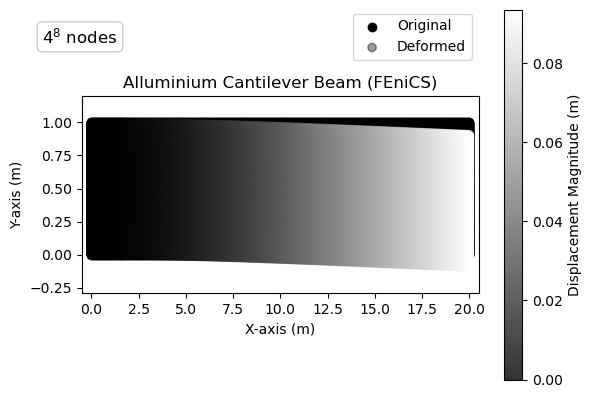}
        \caption{Deformation of the bar obtained with \texttt{FEniCS}.}
        \label{fig:fenics-def}
    \end{subfigure}
    \vspace{1em} 
    \begin{subfigure}[b]{\textwidth}
        \centering
        \includegraphics[width=0.8\textwidth]{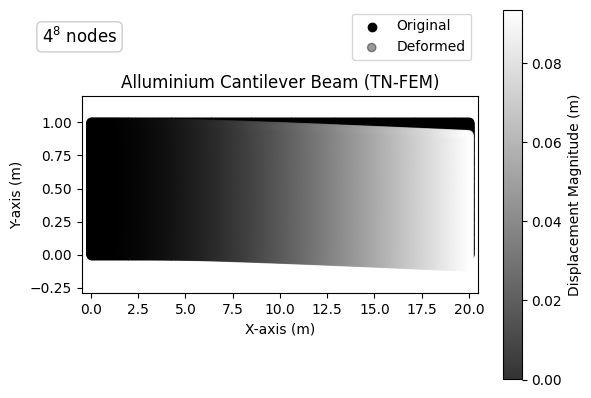}
        \caption{Deformation of the bar obtained with the TN-FEM solver.}
        \label{fig:tnfem-def}
    \end{subfigure}
    \caption{Deformation of the bar. For this plot, we used $4^8$ grid points.}
    \label{fig:deformations}
\end{figure}

\begin{figure}
    \centering
    \includegraphics[width=0.7\linewidth]{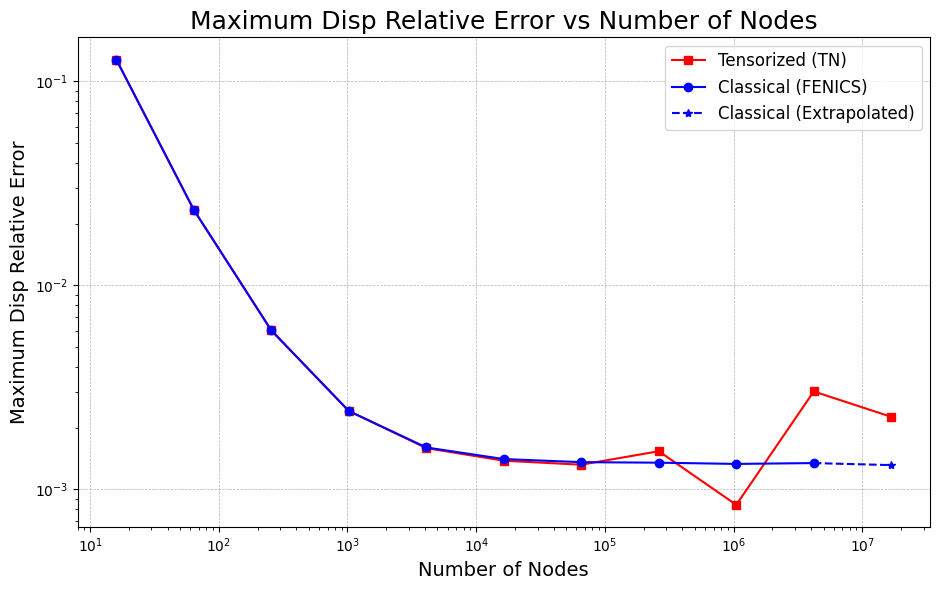}
    \caption{The \texttt{FEniCS} and the TN-FEM solver relative error for the maximum displacement.}
    \label{fig:dipserr}
\end{figure}

\begin{figure}
    \centering
    \includegraphics[width=0.7\linewidth]{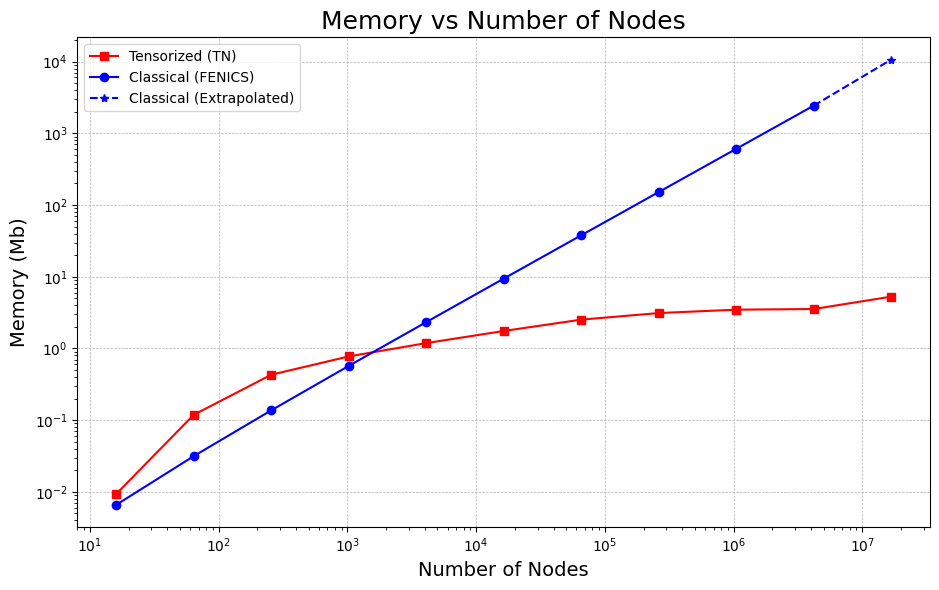}
    \caption{Memory needed to store the stiffness matrix, the force vector and the displacement vector. Both axes are on a logarithmic scale, we observe an exponential compression of the data.}
    \label{fig:memory}
\end{figure}

\begin{figure}
    \centering
    \includegraphics[width=0.7\linewidth]{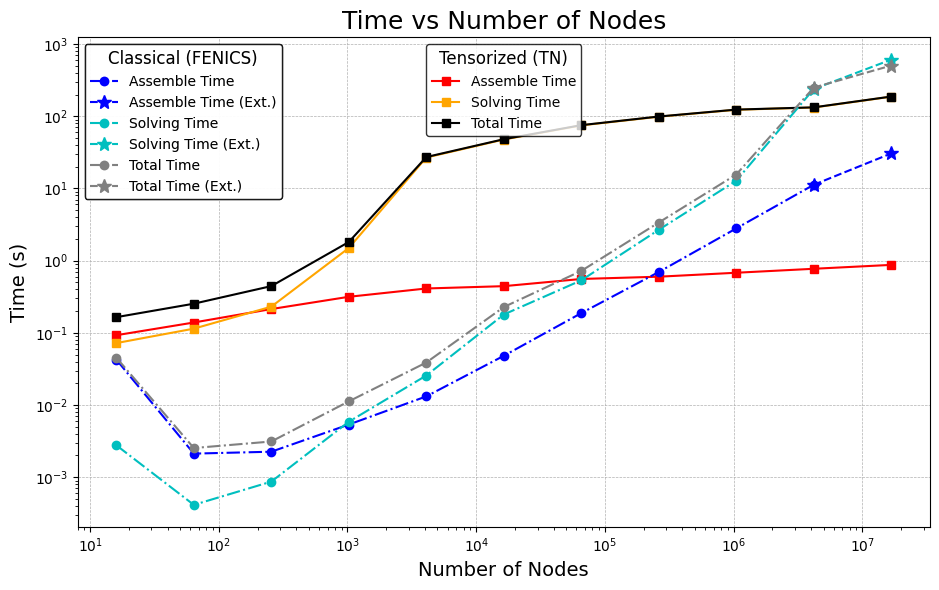}
    \caption{The \texttt{FEniCS} and the TN-FEM solver time required for constructing the FEM system of linear equations, solving time and the total time.}
    \label{fig:time}
\end{figure}

\begin{figure}[H]
    \centering
    \begin{subfigure}[t]{\textwidth}
        \centering
        \includegraphics[width=0.6\textwidth]{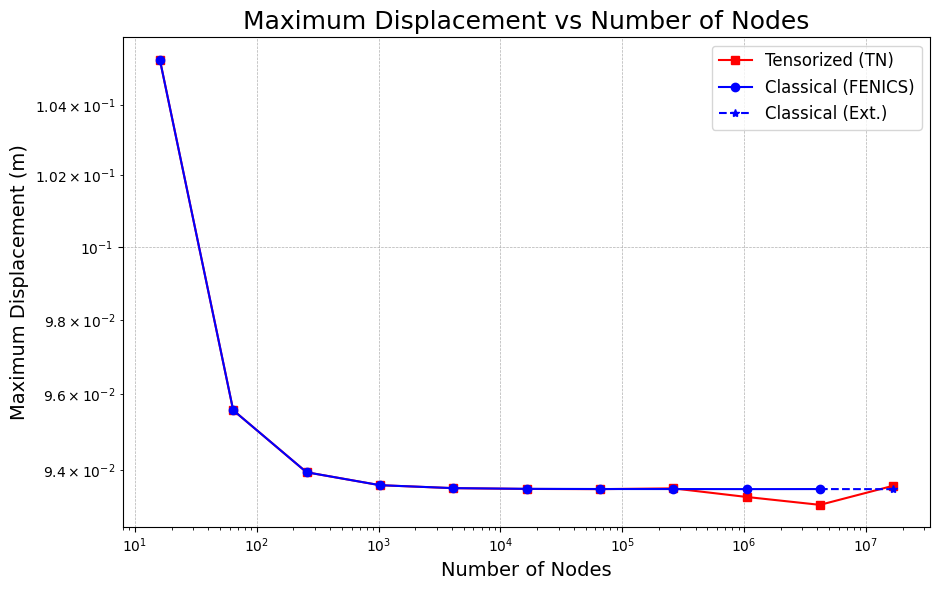}
        \caption{Maximum displacement computed with \texttt{FEniCS} and with our TN-FEM solver.}
        \label{fig:maxd}
    \end{subfigure}
    \vspace{1em} 
    \begin{subfigure}[b]{\textwidth}
        \centering
        \includegraphics[width=0.6\textwidth]{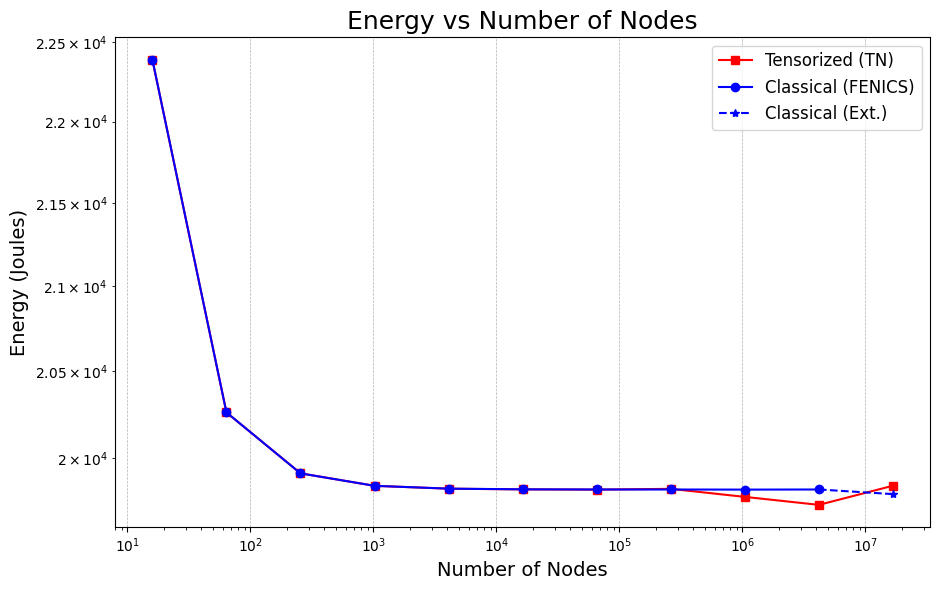}
        \caption{Energy computed with \texttt{FEniCS} and with our TN-FEM solver.}
        \label{fig:energy}
    \end{subfigure}
    \caption{Displacement and energy comparison.}
    \label{fig:obs}
\end{figure}

For comparison, we present the solutions obtained with \texttt{FEniCS}\footnote{A highly optimized open-source classical FEM package: \url{https://fenicsproject.org/}.}, and the results obtained by the TN-solver from 
Algorithm \ref{alg:solver}, see Figures \ref{fig:deformations} -- \ref{fig:obs}. We observe a close agreement between all simulation results.

As a consistency check, we consider the relative error for the the maximum displacement, see Figure \ref{fig:dipserr}. The maximum displacement for the problem that we are considering can be computed analytically as:

\begin{equation}
    |y_{\text{max}}| = \left|\frac{3 \rho g L_{x}^4}{2E L_{y}^2} \right|
\end{equation}
where $g$ is gravity's acceleration, $\rho$ the density of the material, $L_x$ its width, $L_y$ is height and $E$ is the material's Young modulus. The value
$|y_{\text{max}}|$ is shown in
Figure \ref{fig:maxd}, and the relative error is shown in \ref{fig:dipserr} for different discretization sizes.

The main advantage of the TN-solver is its logarithmic scaling with the number of d.o.f., representing an exponential speed-up compared to classical FEM. We illustrate this in Figure \ref{fig:memory} for the memory and in Figure \ref{fig:time} for the time. The computations were performed on a personal MacBook Pro with an Apple M3 Pro chip and \qty{18}{\giga\byte} of RAM. The exact values are less informative than the scaling of the time and space complexity.

The TN-solver will always have some additional overhead depending on the hardware and subtle hyper-parameter tuning of the associated TN-routines. Nonetheless, the logarithmic scaling for the TN-solver is clearly visible in Figures \ref{fig:memory} and \ref{fig:time}. For our ad hoc implementation, the break-even point for the total time to solution is around $10^6$ to $10^7$ d.o.f. For memory consumption, the advantage of the TN approach is even more pronounced.

\section{Conclusion}\label{sec:concl}

Tensor networks are highly sparse multi-linear algebra data structures that can potentially offer exponential savings in both memory and computational time compared to common data structures used in linear algebra packages underlying practically every aspect of numerical computation today. Here, we examined the potential of TNs for simulating material deformations. We demonstrated how to modify the FEM assembly process for the TN format. Our numerical results clearly show the aforementioned exponential advantages in both memory and computational time.

While this project focuses on a simplified example in two dimensions, it establishes a foundation for extending these methods to more complex and industrially relevant applications. The next steps include:
\begin{itemize}
    \item Extension to 3D and more complex domains. This step is a matter of additional implementation; complex domains have already been addressed in \cite{markeeva21}.
    \item Adding TN-preconditioners to ensure stability for very large discretization sizes. Such preconditioners have been previously considered in \cite{bachmayr20}.
    \item Considering non-isotropic materials. This step would modify the stiffness assembly and potentially increase ranks. The overall overhead is likely to increase, but the effect on scaling is unclear at this point.
    \item Considering non-homogeneous materials. This step would require tensorizing the material stiffness as well. The effect on scaling would depend on the TN-ranks of the material tensor.
    \item Considering plasticity and non-linear material behavior. Non-linearities have
    been considered for Navier-Stokes in 2D in \cite{kornev23}. The effect on scaling is unclear at this point.
\end{itemize}

\bibliographystyle{unsrtnat}
\bibliography{bibliography}

\end{document}

%% file: images/drawings/qr_dec.tex
\begin{tikzpicture}

\node at (-1.1, 0.53) {\Large $A$};
\node at (-0.5, 0.5) {\Large $=$};

\draw[thick, rounded corners=0.2cm] (0,0) rectangle (2,1) node[midway] {$A$};

\draw[thick] (1, 1.7) -- (1, 1);
\draw[thick] (1, -0.7) -- (1, 0);
\node at (1.0, 1.95) {j};
\node at (1, -0.9) {i};
\node at (1.3, 1.25) {N};
\node at (1.3, -0.25) {M};

\node at (2.5, 0.5) {\Large $=$};

\draw[thick, rounded corners=0.2cm] (3,0) rectangle (4,1) node[midway] {$Q$};
\draw[thick, rounded corners=0.2cm] (5,0) rectangle (6,1) node[midway] {$R$};

\draw[thick] (3.5, -0.5) -- (3.5, 0);
\node at (3.5, -0.75) {i};
\node at (3.8, -0.25) {M};

\draw[thick] (5.5, 1.5) -- (5.5, 1);
\node at (5.5, 1.75) {j};
\node at (5.8, 1.25) {N};

\draw[thick] (4, 0.5) -- (5, 0.5);
\node at (4.5, 0.75) {r};

\end{tikzpicture}

%% file: images/drawings/TTvec.tex
\begin{tikzpicture}

\def\bigboxwidth{4}
\def\bigboxheight{1}
\def\smallboxsize{1}
\def\spacing{1.5}
\def\cornerradius{5}

\draw[thick, rounded corners=\cornerradius] (0,0) rectangle (\bigboxwidth, \bigboxheight);
\node at (\bigboxwidth/2, \bigboxheight/2) {$V$};

\foreach \i in {0, 1, 2, 3} {
    \pgfmathsetmacro\xpos{\bigboxwidth/8+\i * \bigboxwidth/4}
    \draw[thick] (\xpos, -\spacing/2+0.2) -- (\xpos, 0);
}

\node at (\bigboxwidth/8+0, -\spacing/2 + 0.0) {$i_0$};
\node at (\bigboxwidth/8+\bigboxwidth/4, -\spacing/2 + 0.0) {$i_1$};
\node at (\bigboxwidth/8+\bigboxwidth/2, -\spacing/2 + 0.0) {$i_2$};
\node at (\bigboxwidth/8+3 * \bigboxwidth/4, -\spacing/2 + 0.0) {$i_3$};

\node at (\bigboxwidth + \spacing/4 + 0.4, \bigboxheight/2) {\Huge $=$};

\foreach \i in {0, 1, 2, 3} {
    \pgfmathsetmacro\startx{\bigboxwidth + \spacing + \i * (\smallboxsize + \spacing/2)}
    \pgfmathsetmacro\starty{(\bigboxheight - \smallboxsize) / 2}
    
    \ifnum\i=0
        \draw[thick, rounded corners=\cornerradius] (\startx, \starty) rectangle (\startx + \smallboxsize, \starty + \smallboxsize) node[midway] {$C_0$};
        \node at (\startx + \smallboxsize/2, -\spacing/2 - 0.1) {$i_0$};
        \draw[thick] (\startx + \smallboxsize/2, \starty - \spacing/2+0.2) -- (\startx + \smallboxsize/2, \starty);
    \else
        \pgfmathparse{int(\i)}
        \let\curr\pgfmathresult
        \draw[thick, rounded corners=\cornerradius] (\startx, \starty) rectangle (\startx + \smallboxsize, \starty + \smallboxsize) node[midway] {$C_\curr$};
        \pgfmathsetmacro\labelx{\startx - \spacing/4 + \smallboxsize/2}
        \node at (\startx - \spacing/4, 0.8) {$r_\curr$};
        \node at (\startx + \smallboxsize/2, -\spacing/2 - 0.1) {$i_\curr$};
        \draw[thick] (\startx + \smallboxsize/2, \starty - \spacing/2+0.2) -- (\startx + \smallboxsize/2, \starty);
    \fi
    
    \ifnum\i>0
        \pgfmathsetmacro\prevx{\bigboxwidth + \spacing + (\i-1) * (\smallboxsize + \spacing/2) + \smallboxsize}
        \draw[thick] (\prevx, \starty + \smallboxsize/2) -- (\startx, \starty + \smallboxsize/2);
    \fi
}

\end{tikzpicture}

%% file: images/drawings/TTop.tex
\begin{tikzpicture}

\def\bigboxwidth{4}
\def\bigboxheight{1}
\def\legsize{0.5}
\def\boundsiz{0.5}
\def\smallboxsize{0.8}
\def\spacing{0.4}
\def\cornerradius{2}

\draw[thick, rounded corners=\cornerradius] (0,0) rectangle (\bigboxwidth, \bigboxheight);
\node at (\bigboxwidth/2, \bigboxheight/2) {$A$};

\foreach \x in {0, 1, 2, 3,4} {
    \pgfmathsetmacro\xpos{ \bigboxwidth*(1-0.85)/2+(\x * \bigboxwidth/4)*0.85}
    \draw[thick] (\xpos, \bigboxheight + \legsize) -- (\xpos, \bigboxheight);
    \draw[thick] (\xpos, -\legsize) -- (\xpos, 0);
}

\node at (\bigboxwidth + \spacing*2, \bigboxheight/2) {\Large$=$};

\foreach \i in {1, 2, 3,4,5} {
    \pgfmathsetmacro\startx{\bigboxwidth + \spacing + \i * (\smallboxsize + \spacing)}
    \pgfmathsetmacro\starty{(\bigboxheight - \smallboxsize) / 2}
    
    \draw[thick, rounded corners=\cornerradius] (\startx, \starty) rectangle (\startx + \smallboxsize, \starty + \smallboxsize);
    
    \ifnum\i>1
        \pgfmathsetmacro\prevx{\bigboxwidth + \spacing + (\i-1) * (\smallboxsize + \spacing) + \smallboxsize}
        \draw[thick] (\prevx, \starty + \smallboxsize/2) -- (\startx, \starty + \smallboxsize/2);
    \else
    \fi
    
    \draw[thick] (\startx + \smallboxsize/2, \starty + \smallboxsize + \legsize) -- (\startx + \smallboxsize/2, \starty + \smallboxsize);
    \draw[thick] (\startx + \smallboxsize/2, \starty - \legsize) -- (\startx + \smallboxsize/2, \starty);
}

\end{tikzpicture}

%% file: images/drawings/trapezoid.tex
\begin{tikzpicture}[scale=1.4]

\def\nrows{4}
\def\ncols{4}

\def\ux{1}
\def\uy{0.1}
\def\vx{0.2}
\def\vy{0.8}
\def\wx{-0.1}
\def\wy{-0.06}

\foreach \i in {0,...,\ncols} {
    \foreach \j in {0,...,\nrows} {
        \pgfmathsetmacro\x{\ux * \i + \vx * \j + \wx * \i * \j}
        \pgfmathsetmacro\y{\uy * \i + \vy * \j + \wy * \i * \j}
        
        \fill (\x, \y) circle (1.5pt);
        
        \ifnum\i<\ncols
            \pgfmathsetmacro\nextx{\ux * (\i+1) + \vx * \j + \wx * (\i+1) * \j}
            \pgfmathsetmacro\nexty{\uy * (\i+1) + \vy * \j + \wy * (\i+1) * \j}
            \draw[thick] (\x, \y) -- (\nextx, \nexty);
        \fi
        
        \ifnum\j<\nrows
            \pgfmathsetmacro\nextx{\ux * \i + \vx * (\j+1) + \wx * \i * (\j+1)}
            \pgfmathsetmacro\nexty{\uy * \i + \vy * (\j+1) + \wy * \i * (\j+1)}
            \draw[thick] (\x, \y) -- (\nextx, \nexty);
        \fi
        
        \ifnum\i<\ncols
            \ifnum\j<\nrows
                \pgfmathsetmacro\centerx{\ux * (\i+0.5) + \vx * (\j+0.5) + \wx * (\i+0.5) * (\j+0.5)}
                \pgfmathsetmacro\centery{\uy * (\i+0.5) + \vy * (\j+0.5) + \wy * (\i+0.5) * (\j+0.5)}
                \fill[blue] (\centerx, \centery) circle (1pt);
            \fi
        \fi
    }
}

\draw[->,dashed] (-1, -.1) -> (5, .5) node[below,scale = 0.7] {$x$};
\draw[->,dashed] (-.2, -0.8) -> (1, 4) node[left, scale = 0.7] {$y$};

\draw[->] (-1, 0) -> (5, 0) node[below,scale = 0.7] { $\eta$};
\draw[->] (0, -1) -> (0, 4) node[left, scale = 0.7] {$\xi$};

\node at (-.13, .13) {\small $0$};
\node at (.1, -.13) {\small $0$};
\node at (3, .1) {\small $2^2 \smallminus 1$};
\node at (.3, 2.7) {\small $2^2 \smallminus 1$};

\end{tikzpicture}